\documentclass[fleqn,usenatbib]{mnras}

\usepackage{newtxtext,newtxmath}

\usepackage[T1]{fontenc}

\DeclareRobustCommand{\VAN}[3]{#2}
\let\VANthebibliography\thebibliography
\def\thebibliography{\DeclareRobustCommand{\VAN}[3]{##3}\VANthebibliography}

\usepackage{graphicx}	
\usepackage{amsmath}	
\usepackage{xcolor}
\usepackage[normalem]{ulem}  

\newcommand{\WI}[2]{#1_{\mathrm{#2}}}

\newcommand{\Mtot}{\WI{M}{tot}}
\newcommand{\Mmin}{\WI{M}{min}}
\newcommand{\Mmax}{\WI{M}{max}}
\newcommand{\orb}{\WI{\Omega}{orb}}
\newcommand{\xgr}{\WI{x}{GR}}

\title[Accretion spin-up of the massive component in the neutron star stripping model]{Accretion spin-up of the massive component in the neutron star stripping model for short gamma-ray bursts}

\author[N. Kramarev and A. Yudin]{
Nikita Kramarev,$^{1,2}$\thanks{E-mail: kramarev-nikita@mail.ru}
Andrey Yudin$^{1}$
\\
$^{1}$ National Research Center Kurchatov Institute,
    pl. Kurchatova 1, Moscow, 123182, Russia\\
$^{2}$ Lomonosov Moscow State University, Sternberg Astronomical Institute,
    Universitetsky pr. 13, Moscow, 119234, Russia
}

\date{Accepted XXX. Received YYY; in original form ZZZ}

\pubyear{2023}

\begin{document}
\label{firstpage}
\pagerange{\pageref{firstpage}--\pageref{lastpage}}
\maketitle

\begin{abstract}
In this paper, we use analytical methods to study the last stages of the double neutron star (NS) system evolution. Depending on the initial masses of the components, this evolution can occur either in the framework of the merging scenario or in the NS stripping model. The main new ingredient of this work, compared with previous calculations, is accounting for accretion spin-up of the massive component. This effect leads to a significant decrease in the duration of the stable mass transfer of matter in the stripping mechanism. Within the framework of the Newtonian approximation, we determine the boundary between the merging and stripping scenarios. It is shown that this boundary weakly depends on the total mass of the system and the specific form of the NS equation of state, and is determined mainly by the initial mass ratio of the components. The stripping scenario is realized at $M_2/M_1{\lesssim} 0.8$, so it should make a large contribution to the population of gravitational wave events from NS-NS coalescing binaries which are close to us, and their accompanying short gamma-ray bursts. Nevertheless, the value obtained requires further clarification, taking into account relativistic effects, possible non-conservative mass transfer, etc.
\end{abstract}




\section{Introduction}

Since the discovery of the pulsar PSR B1913+16 \citep{Hulse1975} as a member of a binary neutron star (NS-NS) system, the interest of astrophysicists in such systems has only increased. This is primarily due to the fact that they are natural laboratories to test general relativity (GR) effects. In particular, the analysis of the radio pulses from PSR B1913+16 has indirectly revealed gravitational wave (GW) emission from two NSs spiralling towards each other \citep{Taylor1989}.

The binary neutron star inspirals are also of interest as sources of short gamma-ray bursts (GRBs) \citep{Blinnikov1984,Eichler1989,Narayan1992,Nakar2007}. This process is usually described in the merging model \citep[e.g.][]{Faber2012}, where two NSs approach each other due to the loss of angular momentum to emit GW and finally merge into a single object, a supramassive neutron star (SMNS) or a black hole (BH). However, there is an alternative to this mechanism proposed by \citet{Blinnikov1984}, namely the stripping model. When two NSs spiral inward, the low-mass component first fills its Roche lobe and begins to flow onto its more massive companion. In the end, the low-mass NS (LMNS) reaches the value corresponding to the minimum possible neutron star mass and explodes, actually producing GRB \citep[see also][]{Blinnikov1990}. 

However, for a number of reasons, the stripping model has been neglected for many years. The first reason was connected with the lack of observational data at that time for the NS-NS systems with high mass asymmetry of the components \citep{Thorsett1999}, so necessary for the stripping mechanism. There were also doubts about the existence of stable mass transfer during the stripping of the low-mass component \citep[e.g.][]{Lai1994}. However, the main reason was the weakness of the GRB: the observed short GRBs with known redshifts had, by orders of magnitude, more energy \citep{Fong2015,Lien2016} than that predicted by the stripping model.

Interest in the stripping model was revived on 17 August 2017. After the historical joint detection of the GW event GW170817 and the accompanying GRB170817A, the connection between short GRBs and NS-NS coalescences has been reliably confirmed \citep{Abbott2017a,Abbott2017b}. In our paper \citet{Blinnikov2021}, we have shown that many observational properties of GRB170817A, which turned out to be peculiar compared with other short GRBs, are naturally explained in the context of the stripping model. These properties are the anomalously small total isotropic energy of the GRB, the large ejected mass of the red kilonova \citep{Siegel2019} and its spherical distribution \citep{Sneppen2023}, and the long time delay of $\WI{t}{str}\approx 1.7 \, \text{s}$ between the GW signal peak and the GRB registration \citep{ClarkEardley1977}, corresponding to the duration of the stable mass transfer during the stripping of the LMNS. In subsequent work \citet{Blinnikov2022}, it was shown that the absence of the stable mass transfer observed in some hydrodynamic simulations may be caused by incorrect selection of the initial conditions (see also \citet{Dan2011} for the white dwarf (WD)-WD systems).

In this paper, we examine the properties of the stripping model using the analytical approach of \citet{ClarkEardley1977}. Compared with the previous calculations \citep[see also][]{Jaranowski1992,Imshennik1998,Imshennik2008}, we take into account the effect of accretion spin-up of the massive component obtained earlier in our numerical simulation \citep{Blinnikov2022}. Moreover, we consider the corotation of the LMNS and the tidal spin-down of the massive NS (MNS). This allows us to determine the boundary between the merging and stripping scenarios, as well as the duration of the stable mass transfer $\WI{t}{str}$ (or the stripping time) --- the most important dynamical parameter of the NS stripping model. 

The outline of our paper is as follows. In Section~\ref{ProblemStatement} we describe the problem statement and derive the basic equations that determine the NS-NS system evolution in the stripping model. A discussion of the results is presented in Section~\ref{Results}. Section~\ref{Conclusions} summarizes our main conclusions and outlines the prospects for further progress. 

\section{Problem statement}
\label{ProblemStatement}

\begin{figure}
		\includegraphics[width=\columnwidth]{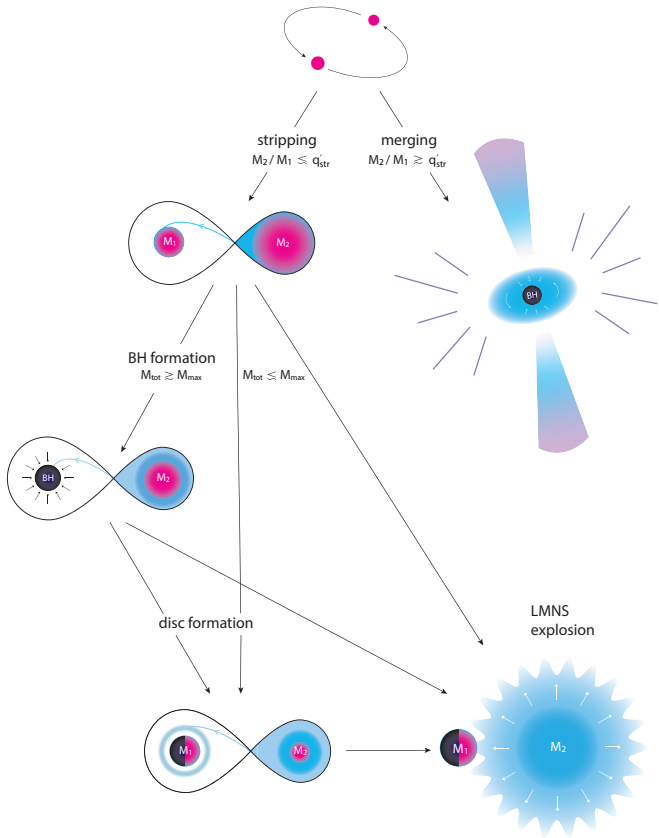}
	\caption{The last stages of the double NS system evolution depending on the initial masses of the components.} \label{Fig.main}
\end{figure}

Let us consider the last stages of the double NS system evolution in more detail, referring to Fig.~\ref{Fig.main}. Two NSs with approximately equal masses approach each other, due to the loss of the total angular momentum of the system $\WI{J}{tot}$ to emit GW, and finally merge (indicated by the arrow to the right). If the binary system has a high enough mass asymmetry of its components (arrow to the left), the LMNS ($M_2$ in the figure) first fills its Roche lobe and begins to flow onto the MNS ($M_1$) through the inner Lagrangian point $L_1$. During accretion, the asymmetry of the system only grows and the components recede each other. The mass transfer lasts on a relatively long time scale, determined by the rate of the angular momentum loss radiated away by GW. At the same time, the orbital angular momentum of the system $\WI{J}{orb}$ is partially transferred to the total rotational (or spin) angular momentum of the MNS and the accretion disc $J_1$. The disc forms when the minimum distance $\WI{R}{m}$ of approach of the stream to the centre of the accretor is larger than the equatorial radius of the MNS $R_1$ \citep{LubowShu1975,Kramarev2022}. Due to the tidal spin-down effect, some part of the rotational angular momentum $J_1$ of the MNS can be transferred back to the orbital angular momentum $\WI{J}{orb}$. By increasing the distance between the components, the LMNS, being in the corotation \citep{Yudin2020}, also spins down and transfers a part of its rotational angular momentum $J_2$ to the orbital one. If the total mass of the system $\Mtot$ is greater than the maximum NS mass $\Mmax$ (see Appendix~\ref{AppendixB}), then at some point the MNS collapses into the BH. In any case, the result of the stripping mechanism is the same (see Fig.~\ref{Fig.main}): once the donor reaches some very low mass value ($\sim 0.2 \WI{M}{\odot}$), the stability of the mass transfer is lost and the remnant $M_2$ is absorbed by $M_1$ on the hydrodynamic time scale. When the LMNS reaches the minimum possible NS mass $\Mmin\sim 0.1 \WI{M}{\odot}$, it loses hydrodynamic stability and explodes.

\subsection{General Assumptions}

Let us move to the mathematical description of the problem statement. As discussed above, the total angular momentum of the system $\WI{J}{tot}$ consists of the orbital angular momentum $J_{\mathrm{orb}}$ and the rotational angular momenta of the components $J_1$ and $J_2$. It is lost due to the GW emission and the magneto-dipole radiation of the rotating MNS. The magneto-dipole losses, as we show in Appendix~\ref{AppendixA}, can be neglected, so the following equality takes place \citep[see also][]{Kremer2015}:
\begin{equation}
\dot{J}_{\mathrm{GW}}=\dot{J}_{\mathrm{orb}}+\dot{J}_1+\dot{J}_2.\label{J_main}
\end{equation}
According to \citet{Kowalska2011}, the orbital eccentricity of most NS-NS systems at the final stages is negligibly small, which, in particular, corresponds to the results of GW170817 and GW190425 data processing \citep{Lenon2020}. Therefore, we consider the binary to be in a circular Keplerian orbit with the orbital rotational frequency
\begin{equation}
\orb=\sqrt{\frac{G\Mtot}{a^3}},\label{O_orb}
\end{equation}
where $a$ is the orbital separation, $\Mtot=M_1+M_2$ is the total mass of the components. The mass transfer is also assumed to be conservative. The orbital angular momentum of the system is defined by the formula 
\begin{equation}
J_{\mathrm{orb}}=\frac{M_1 M_2}{\Mtot}a^2 \orb.\label{J_orb}
\end{equation}
The rate of the orbital angular momentum loss to emit GW is given by \citep[e.g.][]{Paczynski1967}:
\begin{equation}
\dot{J}_{\mathrm{GW}}=-\frac{32}{5}\frac{G}{c^5}\frac{M_{1}^2 M_{2}^2}{\Mtot^2}a^4 \orb^5.\label{J_gw}
\end{equation}

In addition to the definition of $J_1$ and $J_2$, our main equation (\ref{J_main}) must be supplemented by another equation. Prior to the beginning of the mass transfer, this is the condition of constancy of the component masses. Then the low-mass component fills its Roche lobe and the stable mass transfer begins. The closure equation in this case is $R_2=R_{\mathrm{R}}$, where the equatorial radius of the LMNS, $R_2=R_2(M_2,J_2)$, is determined by the NS equation of state (EoS), as well as its spin angular momentum (see Appendix~\ref{AppendixB}). The effective radius of the Roche lobe is parameterized in the standard way \citep{Eggleton1983}:
\begin{equation}
R_{\mathrm{R}}=a f(q'), \; f(q')=\frac{0.49 (q')^{2/3}}{0.6(q')^{2/3}+\ln\big[1+(q')^{1/3}\big]}, \label{R2_Roche}  
\end{equation}
where $q'=M_2/M_1$. Note that we will concurrently use relation $q=M_2/\Mtot$, the ratio of the donor mass to the total mass of the system.

\subsection{Accounting for accretion spin-up of the massive component}

We assume that before the mass-transfer process, both NSs are synchronized due to the tidal effects and corotate, i.e. $\WI{J}{1,2}=I_{1,2}\orb$, where $I_{1,2}$ are the moments of inertia of the accretor and donor. After the beginning of the stable mass transfer, the rotational angular momentum of the massive component changes due to the effects of accretion spin-up and tidal spin-down:
\begin{equation}
\dot{J}_1=\dot{J}_\mathrm{acc}+\dot{J}_\mathrm{tid}.\label{J1_dot}
\end{equation}
For the $\dot{J}_\mathrm{acc}$ term we use the following parameterization:
\begin{equation}
\dot{J}_\mathrm{acc}=-\dot{M}_2 \mathfrak{j}(q,r_1)a^2\orb,\label{J1_acc}
\end{equation}
where $\mathfrak{j}$ is the specific angular momentum of accreting matter in orbital units, $r_1$ is the dimensionless stopping radius (see formula (\ref{r_stop}) below). As discussed above, during the stripping of the low-mass component, two modes of accretion can take place \citep{LubowShu1975}. In the first case, the accretion stream hits the surface of the accretor with the equatorial radius $R_1$ and the so-called direct impact accretion takes place. If the minimum distance $\WI{R}{m}$ for which the stream approaches the massive component is larger than the equatorial radius of the accretor $R_1$, an accretion disc with an outer radius of $\WI{R}{d}$ is formed. The corresponding stopping radius for each mode is equal to
\begin{equation}
r_1 = 
\begin{cases}
R_1/a, \; R_1\geqslant R_{\mathrm{m}}, \\
R_{\mathrm{d}}/a, \; R_1<R_{\mathrm{m}}.
\end{cases} \label{r_stop}
\end{equation}
The detailed calculation and approximation of $\mathfrak{j}(q,r_1)$, as well as $R_{\mathrm{m}}(q)$ and $R_{\mathrm{d}}(q)$ are described in our previous paper \citet{Kramarev2022}. We only emphasize that all our approximations have accuracy better than 1\%, but are obtained in the framework of the Newtonian limit and therefore require further refinement to take into account the GR effects.

\subsection{Accounting for tidal spin-down of the MNS}

Nowadays, several parametric approaches are known to account for the tidal spin-down effect in analytical calculations of the binary system evolution \citep[e.g.][]{Zahn1977,Eggleton1998}. However, the large ambiguity of the input parameters leads to a significant (by orders of magnitude) change in the contribution of this effect \citep[see also][]{Kushnir2017}. To qualitatively account for the tidal spin-down of the MNS, we first consider two limiting cases. In one case, the massive component spins up during accretion according to the formula (\ref{J1_acc}), and there is no tidal spin-down, i.e. $\dot{J}_\mathrm{tid}=0$. In the other limit, the MNS, like the LMNS, rotates synchronously both before and after the beginning of the mass transfer, i.e. $\WI{J}{1}=I_{1}\orb$. Accretion spin-up of the massive component is completely absent in this case. A comparison of calculations is given in Fig.~\ref{Fig.t_str} in Section~\ref{Results}. 

To quantitatively account for the tidal spin-down effect, we exploit the approach used by \citet{Marsh2004,Gokhale2007,Kremer2015} to calculate the evolution of WD-WD binaries. In our case, the term expressing tidal spin-down of the MNS can be parameterized as
\begin{equation}
\dot{J}_\mathrm{tid}=\frac{I_{1}}{\WI{\tau}{syn}}(\orb-\Omega_1), \label{J1_tid}
\end{equation}
where $\Omega_1=J_1/I_1$ is the massive component spin, $\WI{\tau}{syn}$ is its characteristic tidal synchronization timescale. Although the value of $\WI{\tau}{syn}$ can vary within wide limits, we will show below (see Subsection~\ref{SubSecResultTidal}) that in this approach the tidal spin-down effect appears to be secondary to the accretion spin-up. 

\subsection{Corotation of the LMNS}

As was shown in \citet{Yudin2020}, even if there is an initial angular momentum, the LMNS loses it quickly enough during the mass transfer. Therefore, we  assume that the low-mass component always rotates synchronously both before and after the beginning of accretion, i.e. $\WI{J}{2}=I_2(M_2,J_2)\orb$. So the third term on the right-hand side of equation (\ref{J_main}) can be written as
\begin{equation}
\dot{J}_2=\beta_2\bigg[I_2 \dot{\Omega}_{\mathrm{orb}}+\Omega_{\mathrm{orb}}\dot{M}_2 \left(\frac{\partial I_2}{\partial M_2}\right)_{\!J_2}\!\!\bigg],\label{J2_dot}
\end{equation}
where we introduce the notation $\beta_2=\Big[1{-}\Omega_{\mathrm{orb}}\left(\frac{\partial I_2}{\partial J_2}\right)_{\!M_2}\!\!\Big]^{-1}$ for compactness. The calculation procedure of the moment of inertia $I$ and the equatorial radius $R$ of the rotating NS as functions of $M$ and $J$ is described in Appendix~\ref{AppendixB}.

\subsection{The stability criterion for the mass transfer}

For stable mass transfer, the size of the Roche lobe must grow faster than the radius of the low-mass component. In accordance with (\ref{J_main})-(\ref{J2_dot}), the stability criterion can be written as\footnote{For simplicity, here we have neglected the tidal spin-down effect and also assumed $R_{\mathrm{R}}=R_2(M_2)$.}
\begin{equation}
\frac{d \ln R_2}{d \ln M_2}\geqslant \frac{d \ln f}{d \ln q }- 2\frac{1{-}2q{-}\mathfrak{j}(q,r_1){+}\frac{\beta_2}{ a^2}\left(\frac{\partial I_2}{\partial M_2}\right)_{\!J_2}}{1{-}q{-}\frac{3\beta_2 I_2}{ M_2 a^2}}. \label{stab}
\end{equation}
Analysis of this formula shows that the stability of the mass-transfer process is determined mainly by the NS mass-radius relation in the low-mass range, and by the formula for the specific angular momentum of matter going to spin up the accretor.

The derived criterion for the stability of the mass transfer determines the boundary between the merging and stripping scenarios, as well as the duration of the stable mass transfer $\WI{t}{str}$ --- the most important dynamical parameter of the stripping scenario, corresponding to the time delay between the GW signal peak and the GRB detection \citep{Blinnikov2021,Blinnikov2022}.

\subsection{The gravitational wave and neutrino luminosity}
\label{Neutrino}

In the light of the GW170817 and GW190425 registration and the further search for the NS-NS inspirals during the fourth observing run (O4) of the LIGO-Virgo-KAGRA GW detectors \citep{Colombo2022}, we are also interested in the GW luminosity in the stripping mechanism. For two point masses moving in circular orbits the GW luminosity is defined by the well-known formula \citep[e.g.][]{Landau2}:
\begin{equation}
L_{\mathrm{GW}}=\frac{32}{5}\frac{G^4}{c^5}\frac{\Mtot^{5} q^{2} (1{-}q)^{2}}{a^5}. \label{L_gw}
\end{equation}

With the beginning of the stable mass transfer, some part of the accreting matter energy is radiated, following \citet{ClarkEardley1977}, in the neutrino channel. We estimate the neutrino luminosity using the formula
\begin{equation}
L_{\mathrm{\nu}}=\frac{G M_1 \dot{M_1}}{\WI{R}{\nu}}, \label{L_nu}
\end{equation}
where $\WI{R}{\nu}=r_1 a$ is the equatorial radius of the MNS or the radius of the accretion disc (after disc formation). In the case of direct accretion onto the BH, the neutrino luminosity is obviously equal to zero. We note once again that the formula (\ref{L_nu}) is an estimate. Moreover, some part of the energy is also radiated in the electromagnetic channel, which is important in the context of the search for precursor activity of GRBs \citep{Koshut1995, Mereminskiy2022} and the problem of non-conservative mass transfer.

\subsection{Consideration of the GR effects}
\label{GR}

The contribution of the GR effects to our problem can be essential in two hypostases. Firstly, it influences the system parameters: the orbital rotational frequency \citep{Thorne1985} and the effective radius of the Roche lobe of the low-mass component \citep{Ratkovic2005}. From this point of view, the impact of the relativistic effects is significant only at the moment of the closest approach of the components $a\approx40\,\;\mathrm{km}$, when the stripping of matter from the surface of the LMNS begins. For a total mass value of about $\Mtot=2{-}3M_{\odot}$, we obtain the contribution of the relativistic effects on the order of $2G\Mtot/a c^2 \sim 10\%$. It is clear that with an increase in the orbital separation $a$, the impact of the GR effects on the system parameters  decreases.

On the other hand, as discussed in \citet{Kramarev2022}, the relativistic effects also impact the dynamics of the accretion stream near the surface of the MNS and, consequently, the accretor spin-up process at about 10\%. After the collapse of the MNS into the BH, the GR effects must play a major role in the accretion \citep[e.g.][]{NovikovFrolov}. An accurate calculation of the orbital angular momentum transfer to the spin angular momentum of the BH (or the MNS) during accretion would require us to solve the relativistic three-body problem. But instead, we introduce some \textit{effective radius of the BH} at which the accreting matter \textit{stops} and transfers its angular momentum, defined by the approximation formula $\mathfrak{j}(q,r_1)$, to the BH. As such a \textit{radius} we can choose, for example, the radius of the equatorial innermost stable circular orbit of the BH
\begin{equation}
\WI{R}{ISCO}= \frac{R_g}{2}\left(3+Z_2-\text{sign}(l)[(3-Z_1)(3+Z_1+2Z_2)]^{1/2}\right),\label{r_isco}
\end{equation}
including the non-rotating BH radius $3 R_g$ and the limiting rotation radius $4.5 R_g$, where $R_g=2GM/c^2$ is the Schwarzschild radius of the BH with mass $M$. Following \citet{NovikovFrolov}, we determine some auxiliary functions
\begin{equation}
Z_1=1+\left(1-l^2\right)^{1/3}\left[\left(1+l\right)^{1/3}\! \! \!+\left(1-l\right)^{1/3}\right],\label{Z1}
\end{equation}
\begin{equation}
Z_2=\left(3 l^2+Z_1^2\right)^{1/2},\label{Z2}
\end{equation}
where $l=2J/R_g Mc$ is the specific angular momentum of the BH, in the general case varying within $l \in [-1;+1]$. In our case the BH spin is collinear with the orbital angular momentum (and motion of the accreting matter) so $l$ has a positive sign.
One can also choose as such a \textit{radius} the minimum of the periastrons of all parabolic orbits, which in our notation is written as
\begin{equation}
R_b=R_g\left(1-\frac{l}{2}+\sqrt{1-l}\right).\label{r_b}
\end{equation}

By carrying out a series of NS stripping calculations with different formulae for the effective radius of the massive component after its collapse into the BH, we find that the stripping time $\WI{t}{str}$ --- the main parameter of the NS stripping model --- weakly depends on the particular type of the formula (see Subsection~\ref{SubSecResultGR}). This means that the approach described above is acceptable for our problem. Unless otherwise specified, we will use $3 R_g$ as the effective radius of the BH hereafter by default.

\section{Results}
\label{Results}

\subsection{The boundary between the merging and stripping scenarios}

The derived criterion (\ref{stab}) for the stability of the mass transfer, as discussed above, determines the boundary between the merging and stripping scenarios. Before proceeding to the calculations with the specific NS EoS, let us pay attention to the following circumstance. A distinctive feature of the most modern EoSs, satisfying the observational data \citep[e.g.][]{Lattimer2001,Greif2020,Raaijmakers2020}, is that with NS mass variations from one to two solar masses, the NS radius changes by 1-2 kilometres. It turns out that in the moderate mass range, the NS EoS can be described with good accuracy by the polytrope $P(\rho)=K \rho^{1+1/n}$ with index $n{=}1$, for which strictly $R(M){=}\mathrm{const}$ \citep[e.g.][]{Zeldovich1981}. It follows that $R_1=R_2=a f(q')$. We also make use of the fact that the low-mass component, being in the corotation, spins relatively slowly. Then, in the expression (\ref{J2_dot}) for the derivative of the LMNS spin angular momentum, we can put $\beta_2=1$.

In view of the above, the stability criterion for the mass transfer (\ref{stab}) can be rewritten as
\begin{equation}
0\geqslant \frac{d \ln f}{d \ln q }- 2\frac{1{-}2q{-}\mathfrak{j}(q,f)+I^*_2 f^2}{1{-}q{-}3 I^*_2 f^{2}}, \label{stab_approx}
\end{equation}
where $I^*=I/MR^2$ is the dimensionless moment of inertia (see Appendix~\ref{AppendixB}). It is easy to notice that the resulting approximation (\ref{stab_approx}) is determined by the mass ratio of the components $q$ (or $q'$) and weakly depends on the total mass $\Mtot$. This is what we expect from calculations with the different EoSs in agreement with the exact expression for the stability criterion (\ref{stab}).

\begin{figure}
		\includegraphics[width=\columnwidth]{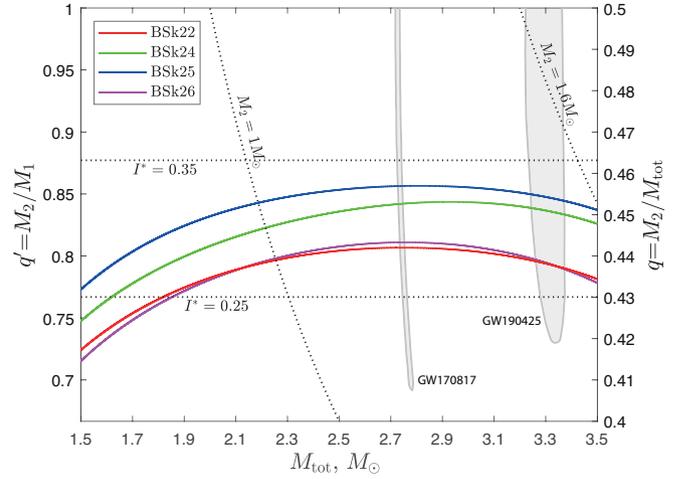}
	\caption{The boundary between the merging and stripping scenarios for the different NS EoS. See text for details.}  \label{Fig.q_M_low}
\end{figure}

Fig.~\ref{Fig.q_M_low} shows the boundary between the merging and stripping scenarios for the BSk22, BSk24, BSk25 and BSk26 EoSs \citep[see][]{Pearson2018}. The horizontal axis represents the total mass of the system, and the vertical axis represents the initial mass ratios of the components $q$ and $q'$. The parameter area $\Mtot{-}q$ above the particular line corresponds to the further merger of the components, and the area below refers to the stripping process of the LMNS. The dotted horizontal lines illustrate the boundary obtained according to the approximate criterion (\ref{stab_approx}) for different values of the dimensionless moment of inertia: $I^*(M_2{=}1 M_{\odot})=0.25$ and $I^*(M_2{=}1.6 M_{\odot})=0.35$. The dotted hyperbolas $M_2=1 M_{\odot}$ and $M_2=1.6 M_{\odot}$ determine the limits of the applicability of our approximation. The slight discrepancy between the prediction of the approximate criterion (\ref{stab_approx}) and the exact numerical calculations is due to the contribution of the logarithmic derivative $d \ln R_2/d \ln M_2$, which is not strictly equal to zero. Despite this, the predictions of the formula (\ref{stab_approx}) are qualitatively correct: the position of the boundary, in fact, weakly depends on the total mass $\Mtot$ and is mainly determined by the mass ratio of the components $q'$. As can be seen from Fig.~\ref{Fig.q_M_low}, the boundary value of the mass ratio is approximately $\WI{q'}{str}(\Mtot)=0.8\pm 0.05$. It is reasonable to compare this value with the constraints on the component masses obtained in the LIGO-Virgo post-processing data analysis for the GW170817 \citep{Abbott2019} and GW190425 \citep{Abbott2020} events. The initial masses for the case of low component spins, corresponding to our corotation case, are represented by the grey regions in Fig.~\ref{Fig.q_M_low}. It can be seen that both detected events could be the result of the NS merging as well as stripping scenarios.

How would the position of the boundary between the scenarios in Fig.~\ref{Fig.q_M_low} change if tidal spin-down of the MNS is taken into account? According to the formula (\ref{J1_tid}), the contribution of this effect to the change of the MNS angular momentum is proportional to the difference between its own and the orbital rotational frequency, $\Omega_1$ and $\WI{\Omega}{orb}$. Therefore, in the case of the MNS corotation, where before the beginning of the mass transfer the frequencies are equal to each other ($\Omega_1=\WI{\Omega}{orb}$), the influence of the mentioned effect must be negligible. However, in the case where the initial rotational frequency of the MNS is more than the orbital one ($\Omega_1>\WI{\Omega}{orb}$), with the beginning of the mass transfer the spin angular momentum of the massive component $J_1$ must effectively transfer to the orbital one $\WI{J}{orb}$. Therefore, in the Newtonian approximation, the case of the initial component corotation illustrated in Fig.~\ref{Fig.q_M_low} gives an absolute lower limit on the boundary $\WI{q'}{str}$ (or $\WI{q}{str}$) between the merging and stripping scenarios. With an increase in the initial rotational frequencies $\Omega_{1,2}$ compared with $\Omega_{\mathrm{orb}}$, the boundary value $\WI{q'}{str}$ between the scenarios also increases, due to the tidal effect as well as the growth of $I_2^{*}$ in the approximate criterion ($\ref{stab_approx}$).

\subsection{The importance of accretion spin-up of the massive component}

\begin{figure*}
		\includegraphics[width=\textwidth]{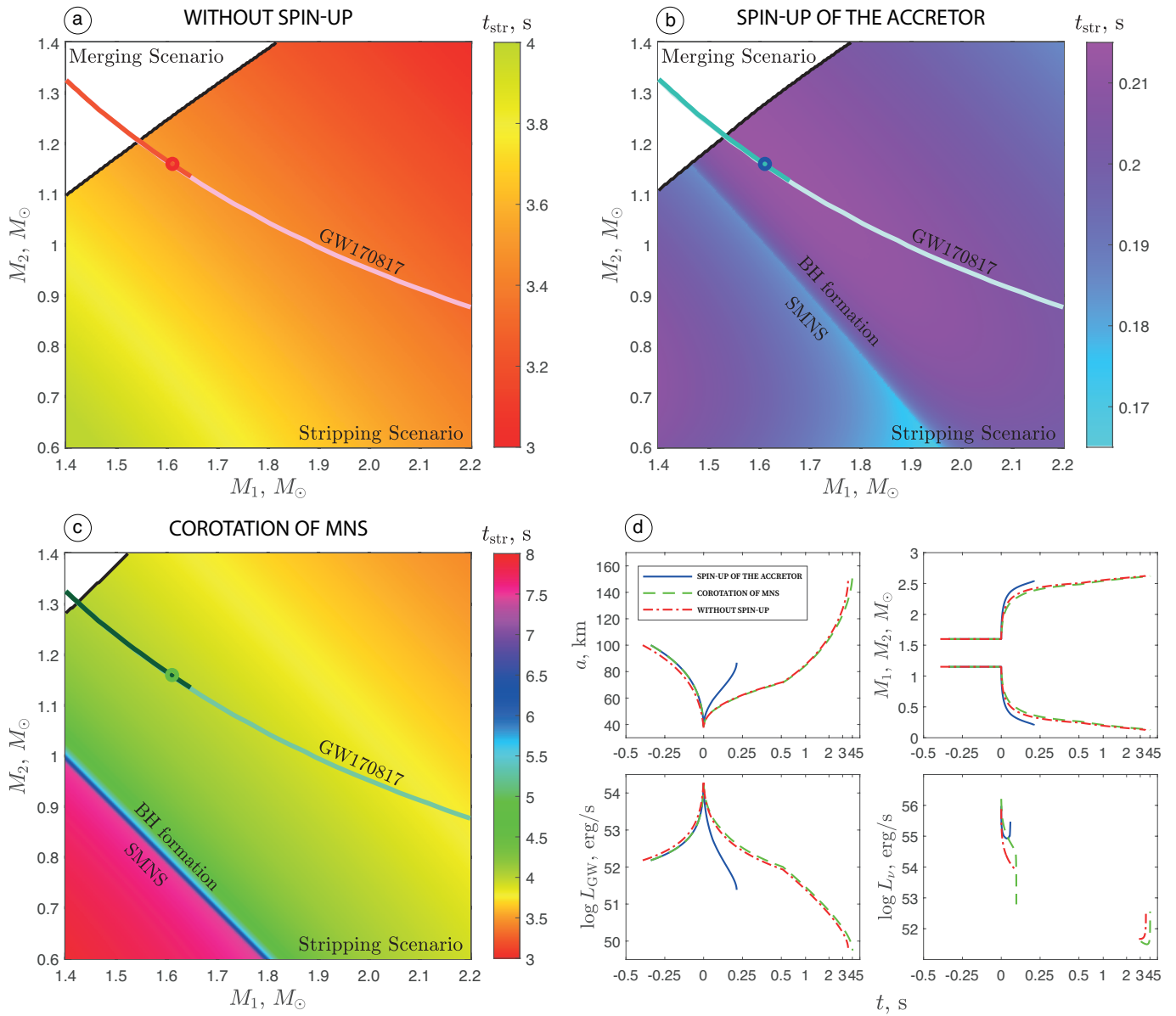} 
        \caption{The stripping time as a function of the initial component masses for the BSk22 EoS. The upper left panel ($a$) illustrates the results of the simplest calculations, made according to formula (\ref{J_simple}), and the upper right panel ($b$) calculations account for the accretion spin-up of the massive component, as well as the corotation of the LMNS, in accordance with (\ref{J_main}). The lower left panel ($c$) calculations take into consideration the tidal spin-down of the MNS, assuming its corotation both before and after the beginning of the mass transfer. The colorbars of panels $a$ and $c$ are consistent with each other. We also add the mass ranges of the GW170817 source: the dark areas correspond to the case of the small initial component spins, and the light areas correspond to the large ones \citep[see][]{Abbott2019}. The blue, red and green circles correspond to the initial masses $M_1=1.6 M_{\odot}$ and $M_2=1.15 M_{\odot}$, for which the evolution of the orbital separation, component masses and luminosities with time are given in the lower right panel ($d$). The drop region of the neutrino luminosity for the green and red lines corresponds to the direct impact accretion onto the BH followed by the formation of an accretion disc. See text for a discussion of all details.} \label{Fig.t_str}
\end{figure*}

Now let us examine the main element of our work --- accounting for accretion spin-up of the massive component. In most previous works \citep{ClarkEardley1977,Jaranowski1992,Imshennik1998,Imshennik2008}, the calculation of the stripping process was performed without taking into account the spin-up of the accretor, i.e. the authors solved a much simpler equation instead of (\ref{J_main}):
\begin{equation}
\dot{J}_{\mathrm{GW}}=\dot{J}_{\mathrm{orb}}. \label{J_simple}
\end{equation}
However, our hydrodynamic simulations \citet{Blinnikov2022} with \texttt{PHANTOM} code \citep{Price2018} showed that the MNS spins up during the stable mass transfer. From general considerations it is clear that accounting for accretion spin-up must lead to a greater loss of the orbital angular momentum of the system and, consequently, to its faster rate of evolution. This is most evident from the comparison of panels $a$ and $b$ in Fig.~\ref{Fig.t_str}. 

Panel $d$ in Fig.~\ref{Fig.t_str} illustrates the evolution of the basic quantities of the NS-NS system with initial masses $M_1=1.6 M_{\odot}$ and $M_2=1.15 M_{\odot}$, corresponding to the mass range of the GW170817 source \citep{Abbott2019}. The zero point in time corresponds to the closest approach of the components and the beginning of the stripping process. The calculations were carried out up to the end of the stable mass transfer according to the criterion (\ref{stab}). The evolution of all quantities at $t \leqslant0.5 \; \mathrm{s}$ is plotted on a linear scale, and at $t>0.5 \; \mathrm{s}$ on a logarithmic scale. The red dash-dotted line corresponds to the calculation in the \citet{ClarkEardley1977} approach, without taking into account accretion spin-up, tidal effects, etc. The blue solid line is the calculation according to the formula (\ref{J_main}), where we account for accretion spin-up of the massive component and corotation of the LMNS. One can see how much accounting for the spin-up effect influences the NS stripping process: the stripping time $\WI{t}{str}$ decreases by an order of magnitude! Hence it follows that accretion spin-up of the massive component must necessarily be taken into account in all future calculations of the NS stripping process.

\subsection{The contribution of the MNS tidal spin-down effect}
\label{SubSecResultTidal}

During tidal spin-down of the MNS, part of its spin angular momentum transfers back to the orbital angular momentum. Therefore, accounting for the tidal spin-down effect must inevitably increase the stripping time. As discussed above, accounting for this effect correctly is quite difficult. So, we first consider the upper limit corresponding to the tidal synchronization of the MNS, both before and after the beginning of the stripping process. For this purpose, in the angular momentum evolution equation (\ref{J_main}) we replace the spin-up formula (\ref{J1_dot}) with (\ref{J2_dot}) (with a corresponding index substitution $2\rightarrow1$). The dependence of the stripping time on the initial masses for the case of the MNS corotation is shown in panel $c$ in Fig.~\ref{Fig.t_str}. Inside the painted part, there are two regions corresponding to the compact object type at the MNS site: SMNS (when $\Mtot < \Mmax \approx 2.4 \WI{M}{\odot}$) or the BH. This circumstance significantly influences the total stripping time, because after the collapse of the SMNS into the BH (BH formation) the tidal spin-down ceases. That is why at $\Mtot \gtrsim \Mmax$ the stripping time $\WI{t}{str}\approx 4 \,\mathrm{s}$, obtained accounting for the tidal spin-down effect, is only slightly longer than the stripping time corresponding to the calculation based on the simplest equation (\ref{J_simple}) (compare panels $a$ and $c$).

The detailed calculation for the initial masses $M_1=1.6 M_{\odot}$ and $M_2=1.15 M_{\odot}$ is shown by the green dashed line in panel $d$ in Fig.~\ref{Fig.t_str}. The green and blue lines coincide until the beginning of the mass transfer, as it should be. As noted above, the total stripping time for the case of the MNS corotation turns out to be slightly longer than the corresponding time obtained from the calculation in the \citet{ClarkEardley1977} approach. This results from the fact that, compared with the case (\ref{J_simple}), with the growth of the separation $a$ between the components of the system, the orbital angular momentum increases due to the decrease in the spin angular momenta of the components, since $J_{1,2} \sim \WI{\Omega}{orb} \sim a^{-3/2}$. It is clear that the true line must lie between the blue and green lines corresponding to the two limiting cases. 

\begin{figure}
				\includegraphics[width=\columnwidth]{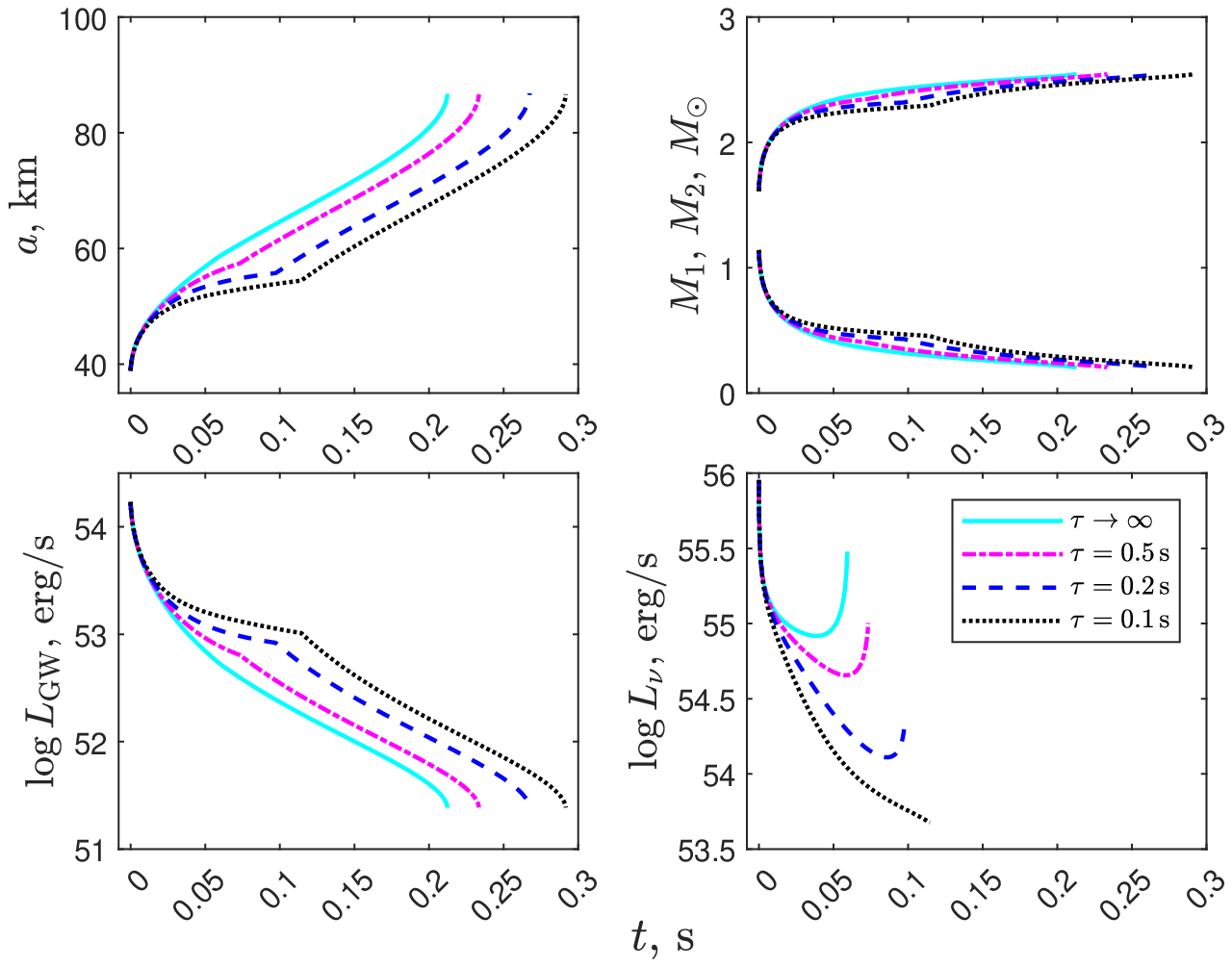}
    \caption{The evolution of the orbital separation, component masses and luminosities for the NS-NS system after the beginning of the mass transfer for the different tidal synchronization timescales $\WI{\tau}{syn}$. The initial component masses are $M_1=1.6 M_{\odot}$ and $M_2=1.15 M_{\odot}$, and the BSk22 EoS was used. The break in the functions $a(t)$, $M_1(t)$, $M_2(t)$ and $\log{\WI{L}{GW}}(t)$ at small $\WI{\tau}{syn}$ corresponds to the derivative discontinuity during the collapse of the MNS into the BH. See text for details.} 
    \label{Fig.stripping_tidal}
\end{figure}

We also quantitatively account for the tidal spin-down effect according to formula (\ref{J1_tid}). The results of calculations for different values of $\WI{\tau}{syn}$ are shown in Fig.~\ref{Fig.stripping_tidal}. It can be seen that even at extremely low values of $\WI{\tau}{syn}$ the stripping time $\WI{t}{str}$ changes by only a few tens of per cent. This is due to the fact that most of the stripping time there is accretion onto the BH that is not affected by the considered effect. Thus, within this approach, the tidal spin-down effect is secondary to the accretion spin-up.

\subsection{The influence of the relativistic effects}
\label{SubSecResultGR}

\begin{figure}
		\includegraphics[width=\columnwidth]{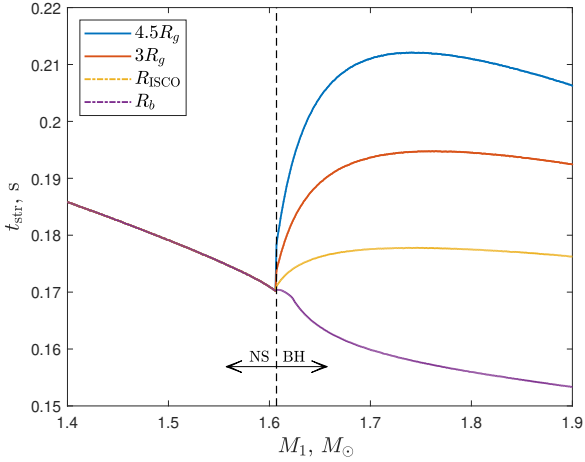}
    \caption{The stripping time as a function of the initial mass of the MNS for the different effective stopping radii of the BH. The initial mass of the low-mass component is $M_2=1 \WI{M}{\odot}$. The area to the right of $M_1\approx 1.6 M_{\odot}$ corresponds to the collapse of the MNS into the BH during the LMNS stripping process. The calculations are performed for the BSk22 EoS.}  \label{Fig.t_m1_m2_1}
\end{figure}

As discussed in Subsection~\ref{GR}, the main manifestation of the relativistic effects is the change of the accreting matter stream dynamics near the surface of the MNS. It is supposed that the GR effects must also be of primary importance when considering the dynamics of the accretion matter falling onto the BH. In order to estimate the influence of the GR effects in calculating the specific angular momentum of accreting matter $\mathfrak{j}(q,r_1)$ going to spin up the BH in accordance with (\ref{J1_acc}), we artificially introduce the effective BH stopping radius $R_1$, which is $3 R_g$, $4.5 R_g$, $\WI{R}{ISCO}$ or $R_b$ (see formulae (\ref{r_isco})-(\ref{r_b})). Fig.~\ref{Fig.t_m1_m2_1} shows how the stripping time --- the main dynamical parameter of our problem --- changes for each radius formula. The small, within hundredths of a second, variations of $\WI{t}{str}$ allow us to argue that the influence of GR effects is not as important as it might seem at first glance. However, it does not eliminate the need to solve the general relativistic problem of the
direct impact accretion onto the MNS or BH.

\section{Conclusions}
\label{Conclusions}

The NS stripping model has experienced a second birth after the joint detection of the GW signal GW170817 and the accompanying GRB170817A \citep{Blinnikov2021,Blinnikov2022}. In the present work we have calculated the long-term evolution of angular momentum in asymmetric NS-NS binaries undergoing the stripping process. We have advanced the analytical approach of \citet{ClarkEardley1977}, mainly accounting for the accretion spin-up of the massive component \citet{Kramarev2022}. It is shown that this effect leads to a significant, more than an order of magnitude, decrease in the stripping time $\WI{t}{str}$ (see Fig.~\ref{Fig.t_str}), which in the paradigm of the stripping mechanism corresponds to the time delay between the loss of the GW signal and the GRB registration ($\WI{t}{str} \approx 1.7 \, \mathrm{s}$ for event GW170817-GRB170817A, according to \citet{Abbott2017b}). Therefore, the spin-up of the massive component must be necessarily taken into account in all subsequent calculations of the LMNS stripping process.

Moreover, we have considered the MNS tidal spin-down and the LMNS corotation. Due to these effects, some part of the rotational angular momentum of components is transferred back to the orbital angular momentum, which leads to only a slight increase in the stripping time. In this context, we have also derived useful approximate formulae to calculate the moment of inertia and equatorial radius of a rotating NS. As shown in Fig.~\ref{Fig_M-R_rotating2}, our formulae produce a good fit to the GR numerical calculations with different EoSs.

\begin{center}
\begin{table}
	\begin{tabular}{p{2.84cm}|p{4.84cm}}
        \hline
        The effect & The impact of considered effect \\
		\hline\hline
		The MNS spin-down via electromagnetic radiation & This effect is \textit{negligible} even when considering rapidly rotating magnetar (see Appendix~\ref{AppendixA}) \\ 
		\hline
		Tidal spin-down of the MNS & It \textit{insignificantly} increases $\WI{t}{str}$ even for extremely low values of $\WI{\tau}{syn}$ (see Fig.~\ref{Fig.stripping_tidal})  \\
		\hline
		The LMNS corotation & Spin of the LMNS is transferred back to the orbital angular momentum during stripping resulting in \textit{small} increase in $\WI{t}{str}$ \\ 
		\hline
		Accretion spin-up of the MNS & This effect leads to a \textit{significant} decrease in $\WI{t}{str}$ (compare panels $a$ and $b$ in Fig.~\ref{Fig.t_str})\\ 
		\hline\hline
        GR correction to the system parameters & We discuss at the beginning of Subsection~\ref{GR} that the orbital rotational frequency and radius of the Roche lobe appear to vary by $10\%$ when mass transfer begins. It can \textit{drastically} change the boundary between scenarios \\
        \hline
		Relativistic modification of direct impact accretion & As it was shown, for example, in Appendix~B of \citet{Kramarev2022}, GR effects can \textit{significantly} change dynamics of accretion near the MNS surface and, respectively, the accretion spin-up process \\ 
		\hline
        Uncertainty in the (cold) NS EoS & According to the criterion (\ref{stab}), the stability of the mass transfer depends on the logarithmic derivative of the function $R(M)$, so $\WI{t}{str}$ must be very sensitive to the NS EoS in the low-mass range \\
        \hline
        Tidal heating of the LMNS crust & Due to the tidal friction of the extended outer LMNS crust during the stripping process, its $R(M)$ relation can \textit{drastically} change in comparison with the cold configuration \\
        \hline
        Non-conservative mass transfer & As we note in Subsection~\ref{Neutrino}, some part of the liberated gravitational potential energy is radiated in the electromagnetic channel. It can lead to a super-Eddington regime of accretion and, correspondingly, non-conservative evolution of a NS-NS system \\
        \hline
	\end{tabular}
        \caption{Upper part: effects considered in this work and their impact on the stripping time $\WI{t}{str}$. Lower part: unaccounted effects and their potential influence.}
        \label{Table.effects}
\end{table}
\end{center}

We summarize different effects and their impact on the stripping process in table~\ref{Table.effects}. In the near future, we intend to examine the influence of the (cold) NS EoS in the low-mass range \citep{Sotani2014}, as we expect it may significantly increase the stripping time. Additionally, we are preparing calculation of the direct impact accretion in various post-Newtonian approaches \citep{Thorne1985,Kaplan2009} to account for the GR effects.

Finally, let us discuss the other important result, related to the determination of the boundary between the NS merging and stripping scenarios. We have shown that this boundary weakly depends on the total mass of the system and the concrete form of the NS EoS (see Fig.~\ref{Fig.q_M_low}) and is determined mainly by the initial mass ratio of the components, which in the considered approach turned out to be $\WI{q'}{str}{=}M_2/M_1{\approx} 0.8$. Based on this high value, about a quarter of the observed galactic NS-NS binaries with known masses \citep{Farrow2019} must finish their evolution in accordance with the stripping model. Therefore, the stripping mechanism must contribute substantially to the total population of close-to-us ($d \lesssim 200 \,\mathrm{Mpc}$) short GRBs, and their corresponding GW sources can be registered during the O4 observing run \citep{Colombo2022}. However, we expect that the subsequent careful accounting for the GR effects, non-conservative mass transfer and other possible factors (see table~\ref{Table.effects}), will decrease the resulting value of the critical binary mass ratio. Even if the fraction of the stripping mechanism in the total population of short GRBs is small, we also expect its large contribution to the production of cosmic heavy elements \citep{Panov2020,Yip2022} due to the relatively large ejected mass ($\Mmin \sim 0.1 \, \WI{M}{\odot}$).

\section*{Acknowledgements}

N.K. is grateful to the RSF 19-12-00229 grant for support. The authors acknowledge Konstantin Manukovsky for calculations of the NS rotating configurations with \texttt{RNS} code. We also thank the reviewer for useful comments.

\section*{Data Availability}
 
Data generated from computations are reported in the body of the paper. Additional data can be made available upon reasonable request.



\bibliographystyle{mnras}
\bibliography{stripping} 




\appendix

\section{The contribution of the MNS electromagnetic emission}
\label{AppendixA}

Let us obtain an upper bound for the effect related to the magneto-dipole angular momentum loss of the MNS. Let $m=B R_1^3/2$ be the magnetic dipole moment of the NS, where $B$ is the magnetic field at the poles. We estimate the angular momentum loss as
\begin{equation}
\dot{J}_{\mathrm{m}}\approx \frac{\dot{E}_{\mathrm{m}}}{\Omega_1},\label{J_m}
\end{equation}
where the energy emission is described by the classical formula \citep[e.g.][]{Landau2}
\begin{equation}
\dot{E}_{\mathrm{m}}=-\frac{2}{3 c^2}m^2 \Omega_1^{4}.\label{E_m}
\end{equation}
The rotational frequency of the NS is bounded above by the Keplerian limit \citep[e.g.][]{Tassoul1978}:
\begin{equation}
\Omega_1 \leq \WI{\Omega}{K} \approx \sqrt{\pi G \bar{\rho}}, \label{Omega_K}
\end{equation}
where $\bar{\rho}=3 M_1/4\pi R_1^{3}$ is the average density of the MNS. 

It is reasonable to compare the obtained expression $\dot{J}_{\mathrm{m}}$ for the magneto-dipole losses with the formula (\ref{J1_acc}) for the angular momentum transferring to the accretion spin-up of the MNS. Taking everywhere $\Omega_1=\sqrt{\pi G \bar{\rho}}$, we can write the upper bound for the momenta relation:
\begin{equation}
\frac{\dot{J}_{\mathrm{m}}}{\dot{J}_{\mathrm{acc}}} \lesssim 0.5 \cdot {10}^{-9} \! \sqrt{\frac{r_1 (1{-}q)^3}{q^2} r_1} \left(\frac{\mathfrak{j}}{0.5}\right)^{-1} \!\! \frac{\WI{t}{str}}{1 \; \mathrm{s}} \left(\frac{B}{10^{12} \; \mathrm{G}}\right)^{2} \!\!\! \frac{R_1}{10 \; \mathrm{km}}.\label{Jm_Jacc}
\end{equation}
Thus, even in the case of the extremely rotating magnetar with $B \sim 10^{15} \; \mathrm{G}$, the magneto-dipole angular momentum loss can be neglected with great accuracy, since $\dot{J}_{\mathrm{m}}/\dot{J}_{\mathrm{acc}} \sim {10}^{-3}$.

\section{The approximate formulae for the moment of inertia and equatorial radius of a rotating NS}
\label{AppendixB}

First of all, let us obtain a consistent description of the $R(M)$ lines of the rotating NSs from the known case of the non-rotating configurations $R_0(M_0)$. It should be taken into account that the case we are interested in --- the spin-up accretion of the NS in the stripping mechanism --- is extremely complicated. In fact, the mass transfer rate is $\dot{M}_1 \sim M_{\odot}/\mathrm{s}$ \citep[e.g.][]{ClarkEardley1977} so the resulting object is likely to be highly hot and unsteady, its rotation should also be non-uniform, etc. So we need a simple and efficient method that will give a good first approximation, and will at least \textit{qualitatively} take into account the basic trends of the phenomenon.

Let us assume that we have a basic parameter in the problem --- the ratio of the spin frequency of the NS $\Omega$ to the maximum rotational (Keplerian) frequency
\begin{equation}
\WI{\Omega}{K}= \sqrt{\frac{GM}{R^3}}.\label{w_k}
\end{equation}
Here we define $\WI{\Omega}{K}$ as the spin frequency at which the equatorial mass shedding begins. The expression (\ref{w_k}) in the general case contains a numerical parameter, which for simplicity we further take as equal to unity. For a given EoS we have a mass–radius relation of the static NS $R_0(M_0)$. The results of the numerical calculations \citep[e.g. Fig.~4 from][]{Martinon2014} motivate us to look for the mass $M$ versus equatorial radius $R$ relation of the rotating NS in the form:
\begin{gather}
M=M_0\left[1+\WI{\alpha}{M}\frac{\Omega^2}{\WI{\Omega}{K}^2}\right]=M_0\left[1+\WI{\alpha}{M}\frac{\Omega^2 R^3}{GM}\right],\label{M_rot_par}\\
R=R_0(M_0)\left[1+\WI{\alpha}{R}\frac{\Omega^2}{\WI{\Omega}{K}^2}\right]=R_0(M_0)\left[1+\WI{\alpha}{R}\frac{\Omega^2 R^3}{GM}\right].\label{R_rot_par}
\end{gather}
These equations are non-linear and make sense only for $\Omega\leq\WI{\Omega}{K}$. The analysis of data on the Keplerian configurations \citep{Lasota1996,Haensel2009,Martinon2014,Li2016} gives the next values of the input parameters: $\WI{\alpha}{M}\approx 0.2$, $\WI{\alpha}{R}\approx 0.4$.

\begin{figure}
		\includegraphics[width=\columnwidth]{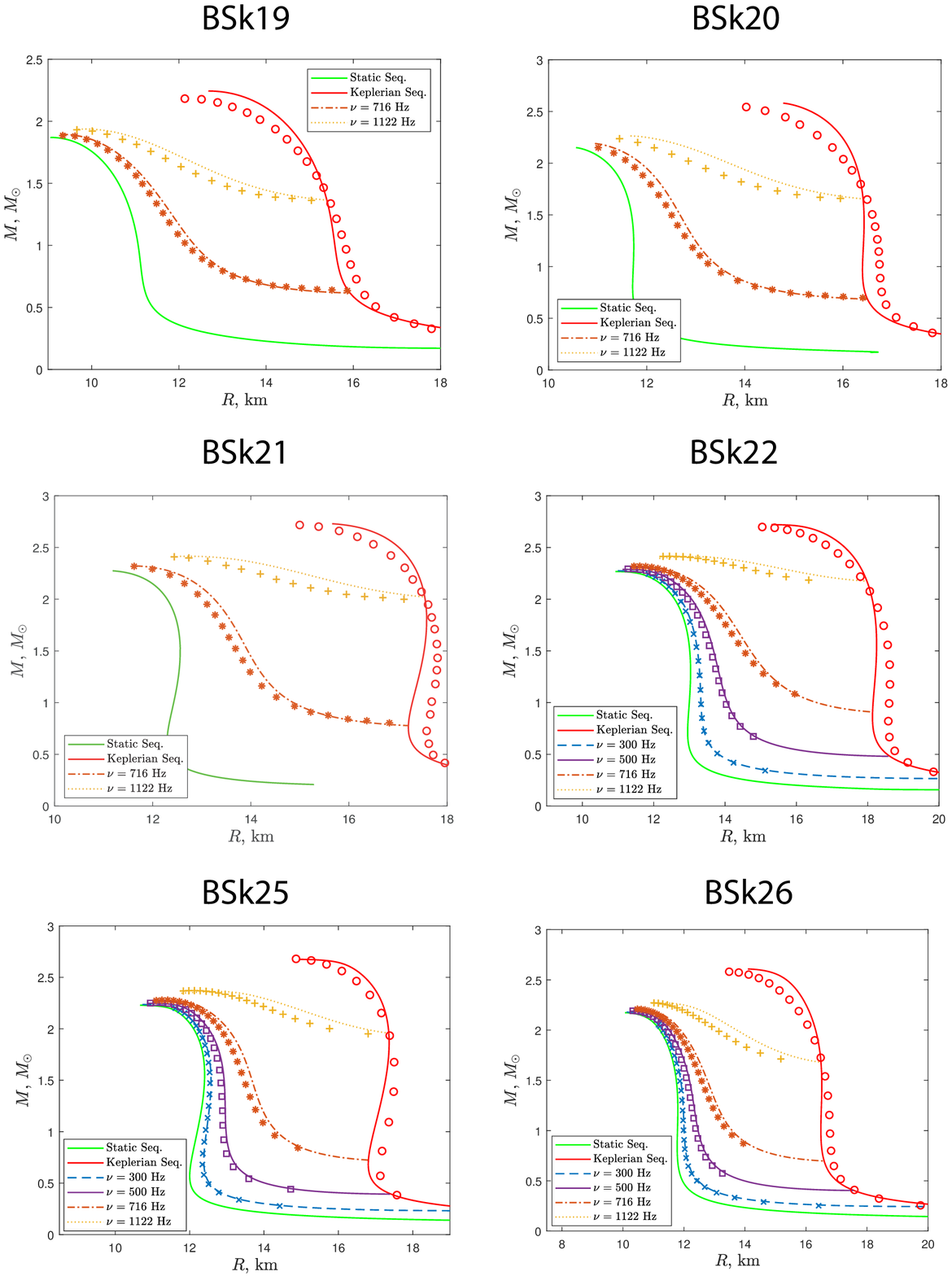}
        \caption{The mass-radius relations of the various rotating configurations for the BSk-type EoSs. The different markers (crosses, asterisks, circles, etc) illustrate the GR numerical calculations and the lines show the results of our approximation (\ref{M_rot_par})-(\ref{R_rot_par}). See text for details.} 
        \label{Fig_M-R_rotating2}
\end{figure}

The mass-radius lines obtained by our approximations (\ref{M_rot_par})-(\ref{R_rot_par}) are shown in Fig.~\ref{Fig_M-R_rotating2}. The NS EoSs are taken from \citet{Fantina2013} and \citet{Pearson2018}. The green line illustrates the static configuration $R_0(M_0)$, and the red one gives the Keplerian configuration with $\Omega=\WI{\Omega}{K}$. The maximum of the red line corresponds to the NS maximum mass $\Mmax$. The different coloured lines show the $M-R$ relations with various frequencies $\nu=\Omega/2\pi$ (in hertz), where $\nu=716\;\mathrm{Hz}$ and $\nu=1122\;\mathrm{Hz}$ correspond to the frequencies of some of the fastest spinning NSs: PSR J1748-2446ad \citep{Hessels2006} and XTE J1739-285 \citep{Kaaret2007}. For comparison, the numerical calculations (different markers) obtained by the GR code \texttt{RNS} \citep{Stergioulas1995,Nozawa1998} are also given. It is seen that our approximation gives not only qualitative, but also a good quantitative agreement with the GR calculations.

Now we determine the moment of inertia of the rotating NS and replace the variable $\Omega$ in the expressions (\ref{M_rot_par}) and (\ref{R_rot_par}) with the new variable $J$ --- the spin angular momentum of the NS. As the dimensionless moment of inertia $I^*=I/MR^2$, we can use, for example, the simplest approximation of \citet{Ravenhall1994}:
\begin{equation}
I^*=\frac{0.21}{1-\xgr},\label{ravenhall}
\end{equation}
where $\xgr=\WI{R}{g}/R=2 G M/ R c^2$. The main feature of our application of this approximation is that we will use it for the rotating stars, substituting into (\ref{ravenhall}) the equatorial radius $R$, which implicitly depends on the mass $M$ and the spin angular momentum $J$ of the star. It is motivated by the fact that at a fixed mass $M$, the moment of inertia $I$ grows with increasing $\Omega$, as it should be, due to a sharp increase of $R$. However, the dimensionless moment of inertia $I^*=I/MR^2$ decreases due to the denominator $1{-}\xgr$, which is also expected from the comparison with the case of the rotating polytropic configurations with index $n=1$ \citep{Cook1994}.

For the particular EoS we can also use more complex approximation formulae for the moment of inertia \citep[e.g.][]{Lattimer2001,Bejger2002}. For the BSk22, BSk24, BSk25 and BSk26 EoSs \citep{Pearson2018}, used in the main part of the article, we derive our own approximation
\begin{equation}
I^* = \frac{a_1 x^{n_1}(1+a_{2}x^{n_2})}{1+a_{3}x^{n_3}}, \label{yudin}
\end{equation}
where $x=M/R$ in $[M_{\odot}/\mathrm{km}]$.
Approximation coefficients (\ref{yudin}) are shown in the table below.
\begin{center}
	\begin{tabular}{|c|c|c|c|c|c|}
		\hline
		$a_1$ & $n_1$ & $a_2$ & $n_2$ & $a_3$ & $n_3$ \\
		\hline
		136.9 & 1.512 & 11.04 & 2.713 & 214.4 & 1.261 \\
		\hline
	\end{tabular}
\end{center}
The presented approximations also satisfy all the properties of the rotating polytropes with index $n=1$ described above: the growth of the moment of inertia $I$ and the decrease in the dimensionless moment of inertia $I^*=I/MR^2$ with increasing $\Omega$ or $J$. Nevertheless, our calculations have shown that the particular type of function $I(M,R)$ has weak influence on the MNS spin-up. Thus, the total stripping time -- our main dynamical parameter --  varies within hundredths of a second for different approximations of the moment of inertia.

Now let us finally formulate our approach. We set one of the formulae for the moment of inertia $I=I(M,R)$ and determine the spin angular momentum $J=I\Omega$. Then the equations (\ref{M_rot_par})-(\ref{R_rot_par}) for the mass and radius of the rotating NS can be rewritten as
\begin{gather}
M=M_0\left[1+\WI{\alpha}{M}\frac{J^2 R^3}{GM I^2}\right],\label{M_rot_J}\\
R=R_0(M_0)\left[1+\WI{\alpha}{R}\frac{J^2 R^3}{GM I^2}\right].\label{R_rot_J}
\end{gather}
We know the mass $M$ and the angular momentum $J$ of the star. Solving the nonlinear equations (\ref{M_rot_J})-(\ref{R_rot_J}) with respect to $M_0$ and $R$, we find the moment of inertia $I(M,R)=I(M,J)$. The static criterion for the maximum mass of an isentropic static configuration discussed by \citet{Bisnovatyi1974} states that it is an extremum (maximum) of the $M$ curve at given $J$ that defines the maximum mass of the star. Exceeding this extremum, the NS collapses into the BH.

In conclusion, we note that the approximation formulae (\ref{M_rot_J})--(\ref{R_rot_J}) that we obtained to calculate $I(M,J)$ and $R(M,J)$ are not universal and require verification on a larger class of the NS EoSs. The approach described above will be refined and developed in another paper.


\bsp	
\label{lastpage}
\end{document}